\documentclass[prd,twocolumn,showpacs,floatfix,amsmath,nofootinbib,amssymb,floatfix]{revtex4}
\usepackage{graphicx,color,dcolumn,booktabs,bm,multirow}
\usepackage{longtable,lscape}
\usepackage{txfonts}
\usepackage{overpic}
\usepackage{amssymb}
\usepackage{indentfirst}
\usepackage{feynmf}   %{feynmp}
\usepackage{slashed}  %for Feynman symbols
\usepackage{cases}
\usepackage{color}
\usepackage{multirow}
\usepackage{epstopdf}
\usepackage{graphicx,color,dcolumn,booktabs,bm}
\usepackage[colorlinks,
            citecolor=blue,
            anchorcolor=red,
            menucolor=red,
            linkcolor=red,
            filecolor=red,
            runcolor=red,
            urlcolor=blue,
            frenchlinks=red]{hyperref}

\def\stat{\mathrm{stat.}}
\def\syst{\mathrm{syst.}}

\begin{document}

%\begin{CJK}{GBK}{}
\title{Where are $\chi_{cJ}(3P)$?}
\author{Dian-Yong Chen$^{1}$}\email{chendy@impcas.ac.cn}
\affiliation{$^1$Department of Physics, Southeast University, Nanjing 210094, People's Republic of China}

\date{\today}
\begin{abstract}
In the present work, we propose $Y(4140)$ as the $\chi_{c1}(3P)$ state by studying the $\chi_{c1} \pi^+ \pi^-$ invariant mass spectrum of the $B\to K \chi_{c1} \pi^+ \pi^-$ process. In the $D\bar{D}$ invariant mass spectrum of the   $B\to K D\bar{D}$ process, we find a new resonance with the mass and width to be $ (4083.0 \pm 5.0) $ and
$ (24.1 \pm 15.4) $ MeV, respectively, which could be a good candidate of the $\chi_{c0}(3P)$ state. The theoretical investigations on the decay behaviors of the $\chi_{cJ}(3P)$ in the present work support the assignments of the $Y(4140)$ and $Y(4080)$ as the $\chi_{c1}(3P)$ and $\chi_{c0}(3P)$ states, respectively. In addition, the $\chi_{c2}(3P)$ state is predicted to be a very narrow state. The results in the present work could be tested by further experiments in the LHCb and forthcoming Belle II.
\end{abstract}

\maketitle
%\end{CJK}

When checking the mass spectrum of the charmonia, one can find the charmonia above 4 GeV are not abundant and our understanding of these states is not comprehensive. In the past decade, a number of the charmonium-like states around 4 GeV have been observed experimentally (see Ref. \cite{Chen:2016qju} for details), which provides us a good opportunity to expand our knowledge of the charmonia spectrum. For the $P$ wave spin-triplet, the ground states have been well established a long time ago, which are $\chi_{c0}(3414)$, $\chi_{c1}(3510)$ and $\chi_{c2}(3556)$ \cite{Agashe:2014kda}. For the first excitation of the $P$ wave state, the $\chi_{c2}(2P)$ state has been confirmed, which was successively discovered in the $\gamma \gamma \to D\bar{D}$ process by Belle and Babar Collaborations \cite{Uehara:2005qd, Aubert:2010ab}. $X(3915)$ observed in the $\gamma \gamma \to J/\psi \omega$ \cite{Lees:2012xs, Uehara:2009tx} and $B\to K J/\psi \omega$ \cite{Abe:2004zs, delAmoSanchez:2010jr} processes is a candidate of $\chi_{c0}(2P)$, and $X(3872)$ first observed in the $B\to K J/\psi \pi^+ \pi^-$ process by the Belle Collaboration in 2003 \cite{Choi:2003ue}, was supported to be the $\chi_{c1}(2P)$ \cite{Danilkin:2010cc, Kalashnikova:2005ui, Zhang:2009bv, Kalashnikova:2009gt, Li:2009zu}.

\begin{figure}[t]
\scalebox{0.55}{\includegraphics{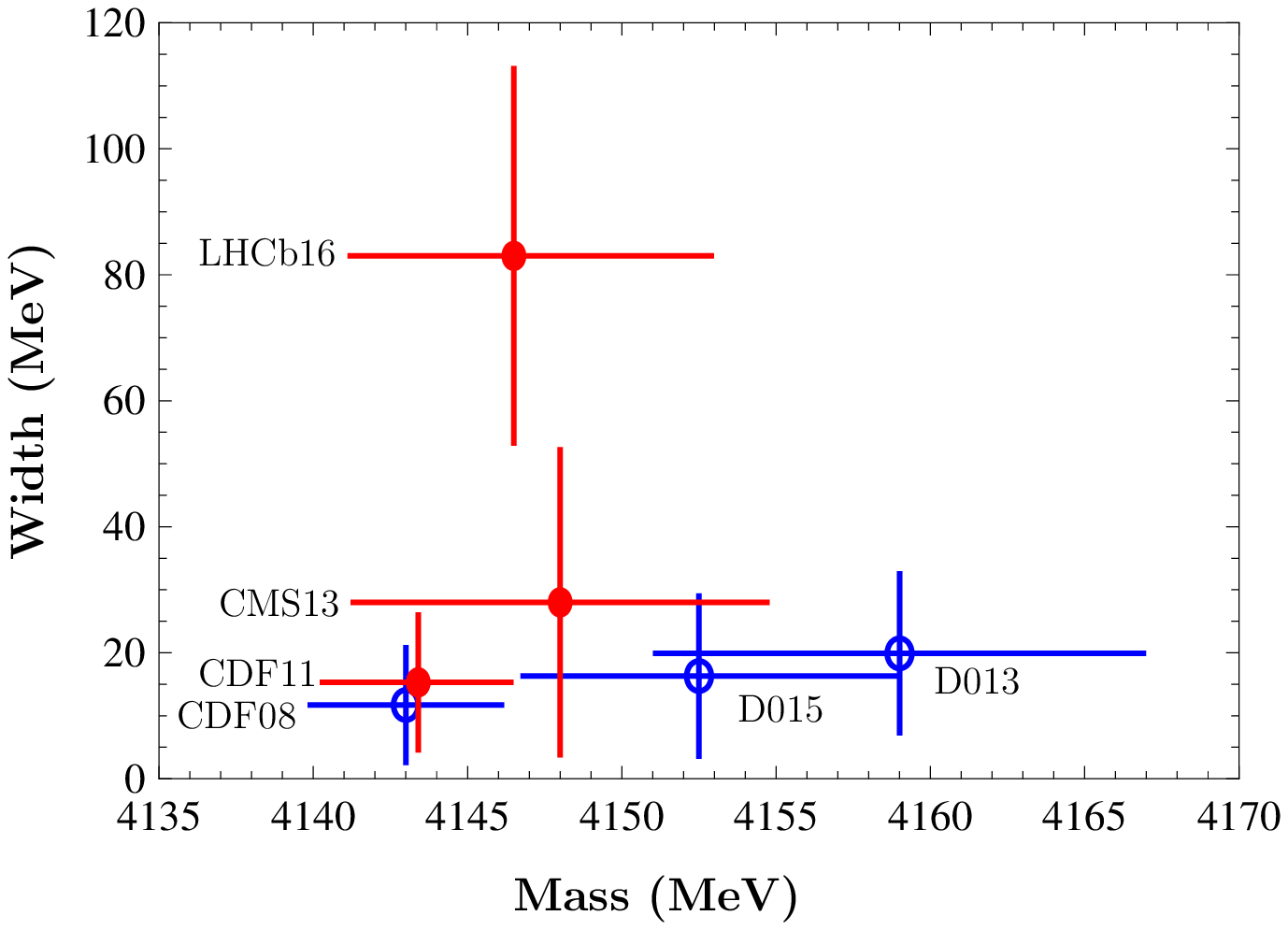}}
\caption{A comparison of the resonance parameters of $Y(4140)$ reported from different measurements \cite{Aaltonen:2009tz, Aaltonen:2011at, Chatrchyan:2013dma, Abazov:2013xda, Abazov:2015sxa, Aaij:2016iza, Aaij:2016nsc}. The filled and open circle indicate the significance of the $Y(4140)$ signals are greater or less than $5 \sigma$, respectively. \label{Fig:expcom}}
\end{figure}

The mass gap of $\chi_{c2}(2P)$ and $\chi_{c2}(1P)$ is about 370 MeV. By adding this mass gap to the mass of  $\chi_{c2}(2P)$, we can roughly estimate that the $\chi_{c2}(3P)$ states should be below 4.3 GeV, since the mass gaps become smaller with the radial quantum number increasing. Thus, the masses of the $\chi_{cJ}(3P)$ state should be smaller than 4.3 GeV, which indicates that the charmonium-like states below 4.3 GeV with the positive $C$ parity could be the candidates of the $\chi_{cJ}(3P)$ states. Very recently, the LHCb Collaboration has confirmed the existence of the $Y(4140)$ and the $J^{PC}$ quantum numbers were determined  to be $1^{++}$ \cite{Aaij:2016iza, Aaij:2016nsc}, so it could be a candidate of the $\chi_{c1}(3P)$ state.

$Y(4140)$ was first observed by the CDF Collaboration in the $J/\psi \phi$ invariant mass spectrum in the exclusive $B^+ \to J/\psi \phi K^+ $ decays with a statistical significance of the signal being $3.8 \sigma$  \cite{Aaltonen:2009tz}. The mass and width of the structure were $4143.0 \pm 2.9 (\stat) \pm  1.2 (\syst)$ MeV and $11.7^{+8.3}_{-5.0} (\stat) \pm 3.7 (\syst)$ MeV, respectively \cite{Aaltonen:2009tz}. Later, the Belle Collaboration measured the cross sections for the $\gamma \gamma \to J/\psi \phi$ and found no evidence of $Y(4140)$ \cite{Shen:2009vs}, which would rule out the $J^{PC}=0^{++},\ 2^{++}$ assignment for the $Y(4140)$. In 2011, the CDF Collaboration reanalyzed the process $B^\pm \to J/\psi \phi K^\pm $ with a larger data sample \cite{Aaltonen:2011at}. The obtained resonance parameters were consistent with the values reported from the CDF previous analysis \cite{Aaltonen:2009tz}, and the statistical significance of the $Y(4140)$ was reported to be greater than $ 5 \sigma$ \cite{Aaltonen:2011at}. However, the previous analyses from the LHCb  and Babar Collaborations did not find the evidence of the $Y(4140)$ state in the $B^+ \to J/\psi \phi K^+ $ and $B^{\pm,0} \to J/\psi \phi K^{\pm,0}$ processes, respectively \cite{Aaij:2012pz, Lees:2014lra}. The CMS Collaboration confirmed the existence of $Y(4140)$ in the $B^{\pm} \to J/\psi \phi K^\pm$ and the significance of the $Y(4140)$ was reported to be greater than $5\sigma$ \cite{Chatrchyan:2013dma}. The D0 Collaboration also observed the signal of $Y(4140)$ in the $J/\psi \phi$ invariant mass spectrum of the $B^+ \to J/\psi \phi K^+$ process and the inclusive $p\bar{p}$ collision process \cite{Abazov:2013xda, Abazov:2015sxa}.

In Fig. \ref{Fig:expcom}, the mass and width of Y(4140) reported from different collaborations are presented \cite{Aaltonen:2009tz, Aaltonen:2011at, Chatrchyan:2013dma, Abazov:2013xda, Abazov:2015sxa, Aaij:2016iza, Aaij:2016nsc}. It should be noticed that the width reported by the LHCb Collaboration is obviously larger than the one from other experimental groups, while the measurements from other three experimental collaborations are in line with each other. Such a discrepancy is interesting and need more experimental efforts from different experimental groups. In addition, the observation of the $Y(4140)$ in other channels could also provide a key of resolving this discrepancy.

\begin{figure}[htb]
\scalebox{0.30}{\includegraphics{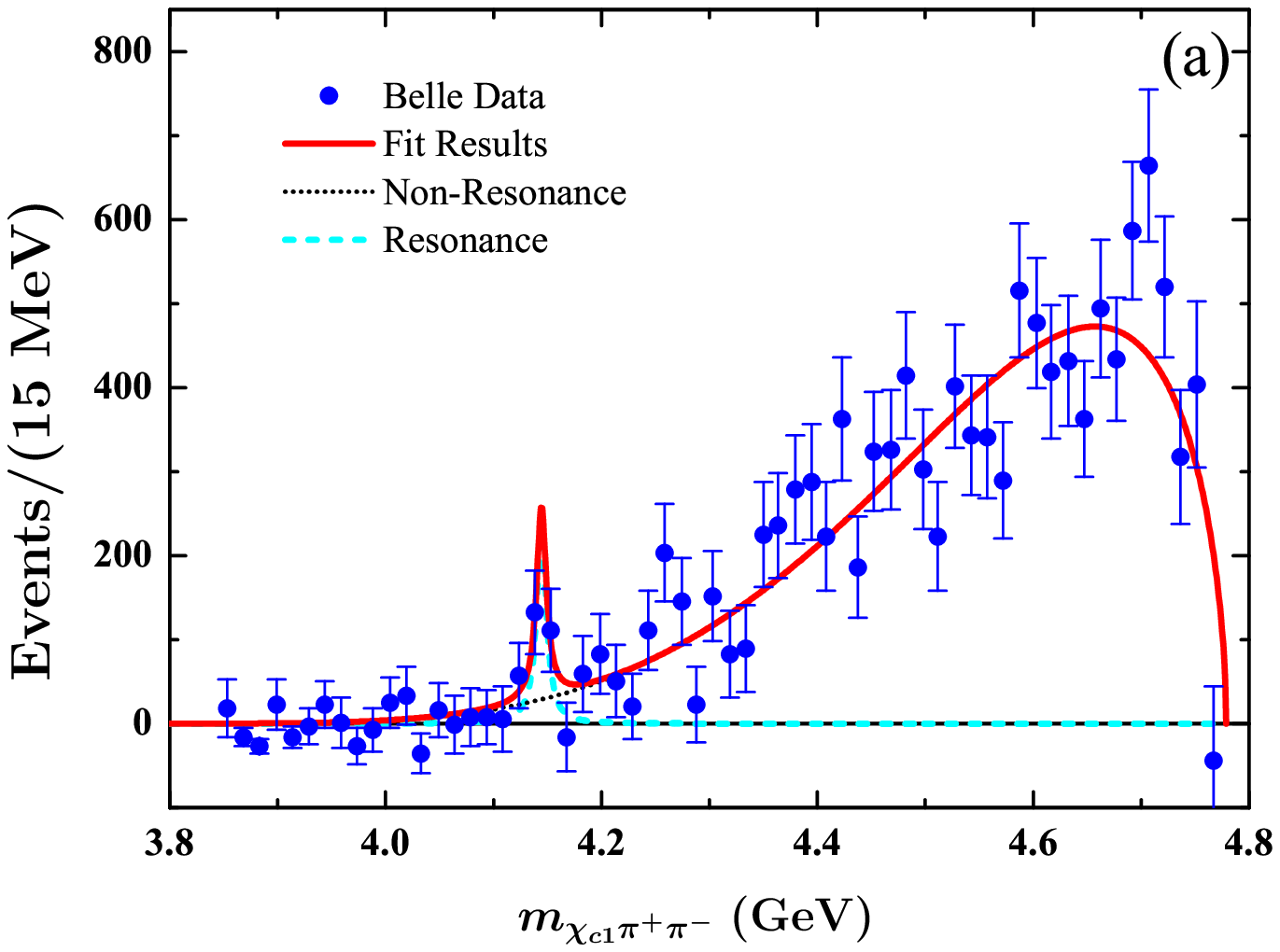}} %\hspace{6mm}
\scalebox{0.30}{\includegraphics{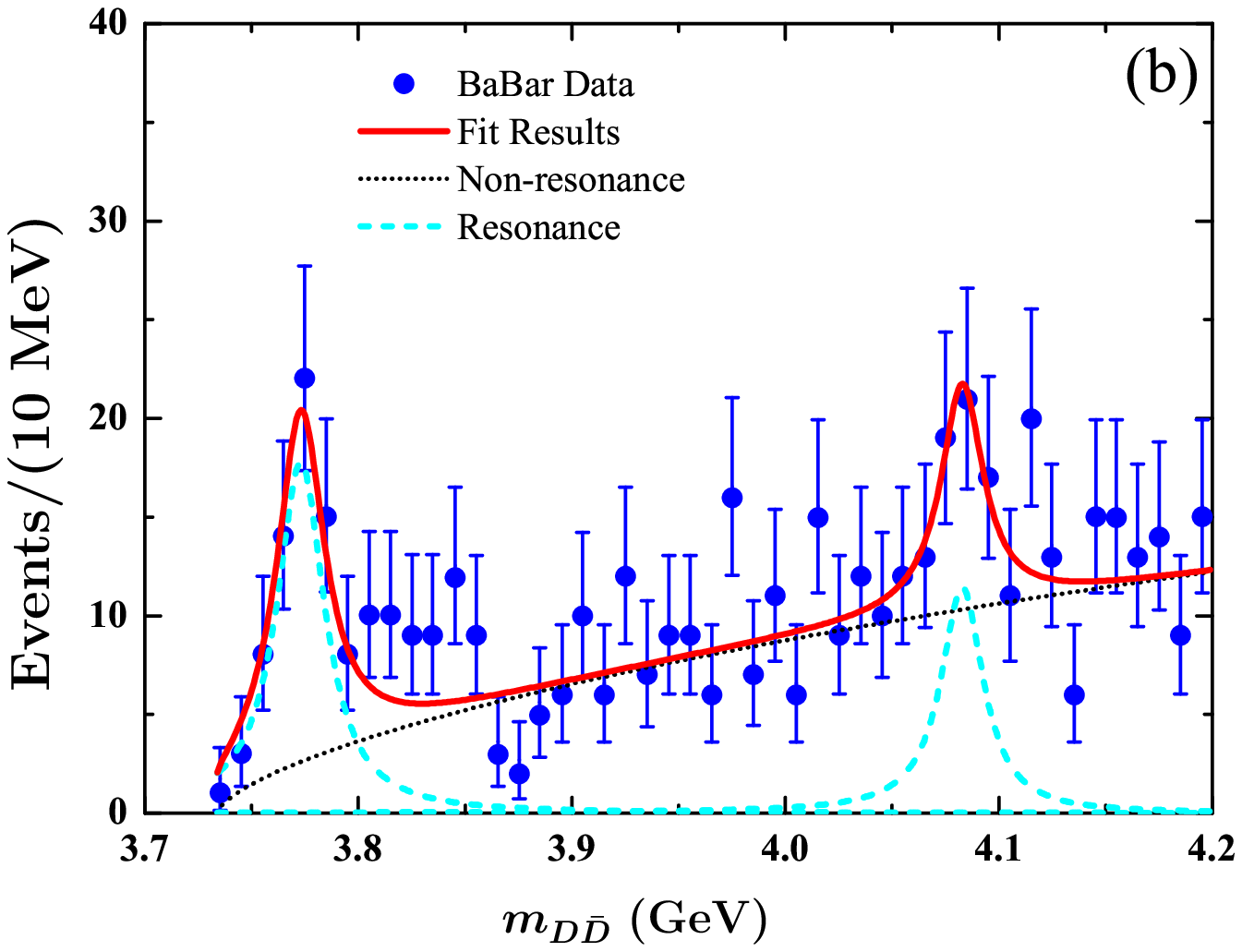}}
\caption{Our fit to the efficiency corrected $\chi_{c1} \pi^+ \pi^-$ invariant mass spectrum of the $B\to K \chi_{c1} \pi^+ \pi^-$ (a) from the Belle Collaboration \cite{Bhardwaj:2015rju} and the $D\bar{D}$ invariant mass spectrum of the $B\to K D\bar{D}$ (b) from the Babar Collaboration \cite{Aubert:2007rva}. \label{Fig:fit}}
\end{figure}

In Ref. \cite{Bhardwaj:2015rju}, the Belle Collaboration reported their inclusive and exclusive measurements of $B$ decays to $\chi_{c1}$ and $\chi_{c2}$. When looking at the $\chi_{c1} \pi^+ \pi^-$ invariant mass spectrum of the $B^+ \to K^+ \chi_{c1} \pi^+ \pi^-$ process, one can find a number of event near 4.1 GeV. Searching the signal of $Y(4140)$ in the $\chi_{c1} \pi^+ \pi^-$ invariant mass spectra could provide us more information on the intrinsic nature on $Y(4140)$. In the present work, we introduce one resonance near 4.1 GeV as well as a non-resonance contributions to fit the experimental data for the $\chi_{c1} \pi^+ \pi^-$ invariant mass distributions reported by the Belle Collaboration \cite{Bhardwaj:2015rju}. The non-resonance contribution is phenomenologically described by
\begin{eqnarray}
\mathcal{A}_{\mathrm{NonR}}  \sim v^h u^p e^{c u},
\end{eqnarray}
where $u=1-m^2/m_{\mathrm{U}}^{2}$ and $ v=m^2/m_{\mathrm{L}}^2-1 $ with $m_{\mathrm{U}}$ and $m_{\mathrm{L}}$ are the upper and lower thresholds of the $\chi_{c1} \pi^+\pi^-$ invariant mass distributions and $m$ is the invariant mass of $\chi_{c1} \pi^+\pi^-$.

As for the resonance contribution, a phase space corrected Breit-Wigner distribution is adopted, which is in the form
\begin{eqnarray}
\mathcal{A}_{\mathrm{BW}}^{R} = \frac{\mathcal{P}(m)}{\mathcal{P}(m_R)} \frac{f_R m_R \Gamma_R}{(m_R^2-m^2)-im_R \Gamma_R}
\end{eqnarray}
where $m_R$ and $\Gamma_R$ are the mass and width of the resonance, respectively. $f_R$ is the coupling constant, which will be treated as a free parameter in the fitting. $\mathcal{P}(m)$ is the phase space of the $B\to K R$, which is
\begin{eqnarray}
\mathcal{P}(m) =\frac{1}{16 \pi} \frac{1}{m_B^3} \lambda(m_B^2,m_K^2,m^2)^{1/2}
\end{eqnarray}
with K$\ddot{\mathrm{a}}$llen  function $\lambda(a,b,c)=a^2+b^2+c^2-2(ab+ac+bc)$.

\begin{table}[t]
\caption{Parameters obtained from fitting to the invariant mass spectra of the $\chi_{c1} \pi^+ \pi^-$ of the $B\to K\chi_{c1} \pi^+ \pi^-$ process and the $D\bar{D}$ of the $B\to K D\bar{D}$ process \cite{Bhardwaj:2015rju, Aubert:2007rva}. \label{Tab:para1}}
\begin{tabular}{cccc}
\toprule[1pt]\toprule[1pt]
$B\to K \chi_{c1} \pi^+ \pi^-$\\
\midrule[1pt]
%Parameter & Value &Parameter & Value  \\
%\midrule[1pt]
$h$ & $2.12 \pm 0.02$ & $p$  &$ 0.29 \pm 0.06$ \\
$c$ & $0.90 \pm 0.18$ & $f_{R}$ & $15.16 \pm 2.92$ \\
$m_{R}$ & $(4144.5 \pm 2.3)$ MeV & $\Gamma_{R}$ & $(11.0 \pm 8.7)$ MeV \\
\midrule[1pt]\midrule[1pt]
$B\to K D\bar{D}$\\
\midrule[1pt]
$h$ & $0.32 \pm 0.09$ & $p$  &$ 0.01 \pm 0.42$ \\
$c$ & $0.32 \pm 0.30$ & $f_{\psi(3770)}$ & $4.21 \pm 0.35$ \\
$m_{R}$ & $(4083.0 \pm 5.0)$ MeV & $\Gamma_{R}$ & $(24.1 \pm 15.4)$ MeV \\
$f_R$ & $3.37 \pm 0.62$ \\
\bottomrule[1pt]\bottomrule[1pt]
\end{tabular}
\end{table}

The efficiency corrected $\chi_{c1} \pi^+ \pi^-$ invariant mass distributions could be described by the incoherent sum of the non-resonance and the resonance contributions. The fitting result is presented in Fig. \ref{Fig:fit}a, in which the individual contributions from the resonance and the non-resonance are also given. The fitting curve could well reproduce the structure near 4.1 GeV. The fitting parameters are listed in Table \ref{Tab:para1}. The resonance parameters of the structure near 4.1 GeV are fitted to be,
\begin{eqnarray}
m&=& (4144.5 \pm 2.3)\ \mathrm{MeV}, \nonumber\\
\Gamma &=& (11.0 \pm 8.7) \mathrm{MeV}, \nonumber
\end{eqnarray}
respectively, which are well consistent with resonance parameters of $Y(4140)$ reported by CDF and CMS Collaborations \cite{Aaltonen:2011at, Chatrchyan:2013dma} and the $J^{PC}$ quantum numbers of this state could be $1^{++}$. Thus, the state in th $\chi_{c1} \pi^+ \pi^-$ invariant mass spectrum could be the same state as $Y(4140)$. In this case, the $Y(4140)$ could couple to both $\chi_{c1} \pi^+ \pi^-$ and $J/\psi \phi$, it is unlikely to be pure $c\bar{c} s\bar{s}$ tetraquark state, molecular state composed from $D_s^{\ast +} D_s^{\ast -}$ or cusp effect of the $D_s^{\ast +} D_s^-+h.c.$ proposed in previous literature \cite{Mahajan:2009pj,Ding:2009vd, Liu:2009ei,  Molina:2009ct, Liu:2009pu, Zhang:2009vs, Albuquerque:2009ak, Zhang:2009st, Wang:2014gwa, Swanson:2014tra, Swanson:2015bsa, Chen:2016ugz, Branz:2009yt, Wang:2009ue, Wang:2009ry, Wang:2016tzr, Stancu:2009ka, Patel:2014vua}. Here, we propose that the $Y(4140)$ could be a $P-$wave charmonium state, i.e., $\chi_{c1}(3P)$, which is the second radial excitation of $\chi_{c1}(3510)$. In such an assignment, $Y(4140)$ could couple to charmed and charmed-strange meson pairs and then couple to the $J/\psi \phi$ or $\chi_{c1} \pi^+ \pi^-$ via charmed-strange or charmed meson loops, which is a typical mechanism working in the light meson transitions between heavy quarkonia \cite{Chen:2011qx, Li:2007au, Zhang:2009kr, Chen:2014ccr, Guo:2009wr, Chen:2014sra}.

In addition, from the present fit to the $\chi_{c1} \pi^+ \pi^-$ invariant mass spectrum, one could find the ratio of $\mathcal{B}[B^+ \to K^+ Y(4140), Y(4140) \to \chi_{c1}\pi^+ \pi^-]$ and $\mathcal{B}[B^+ \to K^+ \chi_{c1} \pi^+ \pi^-]$ to be about $2\%$ when taking the center values of the fitting parameters. The branching ratio $\mathcal{B}[B^+ \to K^+ \chi_{c1} \pi^+ \pi^-]$ was reported to be $(3.74 \pm 0.18 \pm 0.24) \times 10^{-4}$ by the Belle Collaboration \cite{Bhardwaj:2015rju}, thus, one could roughly estimate the branching ratio of the cascade process $B^+ \to K^+ Y(4140), Y(4140) \to \chi_{c1} \pi^+ \pi^-$ to be about $7 \times 10^{-6}$. This branching ratio is of the same order as the $\mathcal{B}[B^+ \to K^+ Y(4140), Y(4140) \to J/\psi \phi]$, which is  $(10 \pm 5) \times 10^{-6}$. This conclusion is consistent with our expectation due to the similarity of these two processes. Furthermore, if the $X(3872)$ is considered as the candidate of the $\chi_{c1}(2P)$, we find the branching ratio $\mathcal{B}[B^+ \to K^+ Y(4140), Y(4140) \to \chi_{c1} \pi^+ \pi^-]$ is also similar to $\mathcal{B}[B^+ \to K X(3872), X(3872)\to \pi^+ \pi^- J/\psi]$ and $\mathcal{B}[B^+ \to K X(3872), X(3872)\to \omega J/\psi]$, which are $(8.6 \pm 0.8)\times 10^{-6}$ and $(6.0 \pm 2.2)\times 10^{-6}$ \cite{Agashe:2014kda}, respectively. Thus, the assignment of the $Y(4140)$ as the $\chi_{c1}(3P)$ state does not conflict with the present experimental data.

If $Y(4140)$ could be assigned as the $\chi_{c1}(3P)$ state, the mass of the  $\chi_{c0}(3P)$ states should be below 4.14 GeV. In Ref. \cite{Aubert:2007rva}, the Babar Collaboration reported their measurement of the $D\bar{D}$ invariant mass spectrum of the $B\to K D\bar{D}$, in which one can find the signal of $\psi(3770)$ and a number of events below 4.1 GeV. In a similar way, we fit the $D\bar{D}$ invariant mass spectrum of the $B\to K D\bar{D}$ process with a non-resonance contributions and two resonances. Here, the resonance parameters of $\psi(3770)$ are fixed to be the PDG average values \cite{Agashe:2014kda} and leave the coupling constant $f_{\psi(3770)}$ as a free parameter. The fitting curve as well as the individual contributions are presented in Fig. \ref{Fig:fit}b, in which the structure near 4.1 GeV is well reproduced. The mass and width of this state are fitted to be
\begin{eqnarray}
m &=& (4083.0 \pm 5.0)\ \mathrm{MeV}\nonumber\\
\Gamma &=& (24.1 \pm 15.4) \ \mathrm{MeV},
\end{eqnarray}
respectively. This state, named $Y(4080)$, could be a good candidate of $\chi_{c0}(3P)$. From the fitting parameters, one can find that $Y(4080)$ is just about 60 MeV below $Y(4140)$. This mass gap is very reasonable compared to the one of the $1P$ state, which is about 100 MeV. In addition, the $\chi_{c0}(3P)$ state couples to the $D\bar{D}$ via an $S-$wave, thus, the observation of $\chi_{c0}(3P)$ state in the $D\bar{D}$ final states is consistent with the expectation.

Similar to the case of $B\to K \chi_{c1}\pi^+ \pi^-$, one can roughly estimate the branching ratio of $B\to K \chi_{c0}(3P), \chi_{c0}(3P)\to D\bar{D}$ from the present fitting results. With the center values of the fitting parameter listed in Table I, the ratio of $\mathcal{B}[B^+ \to K^+ \chi_{c0}(3P) , \chi_{c0}(3P) \to D\bar{D} ]$ and $\mathcal{B}[B^+ \to K^+ \psi(3770) , \psi(3770) \to D\bar{D}]$ is estimated to be 0.6. Thus, branching ratio $\mathcal{B}[B^+ \to K^+ \chi_{c0}(3P) , \chi_{c0}(3P) \to D\bar{D} ]$ is about $1.5 \times 10^{-4}$. Since both $\psi(3770)$ and $\chi_{c0}(3P)$ dominantly decay into $D\bar{D}$, thus, one could conclude that the branching ratio of $B^+\to K^+ \chi_{c0}(3P) $ is about $1.5 \times 10^{-4}$. This branching ratio is of the same order as the one of $B^+ \to K^+ \chi_{c0}(1P) $, which is $1.50^{+0.15}_{-0.14}\times 10^{-4}$. Thus, $Y(4080)$ could be a candidate of the $\chi_{c0}(3P)$ state.

To test the possibility of $Y(4140)$ and $Y(4080)$ as the $\chi_{c1}(3P)$ and $\chi_{c0}(3P)$ states, respectively, we can further check the mass spectrum of charmonia and the decay behavior of the $\chi_{cJ}(3P)$ states.

{\it Mass spectrum:---}
The mass spectrum of the charmonium has been widely investigated in potential model. The Godfrey and Isgur performed a systematic investigation of the spectra of the meson system in a relativistic quark model with the potential of the quark-antiquark in a linear form in the large distance \cite{Godfrey:1985xj}, which is consistent with the Lattice QCD calculation with quenched approximation \cite{Born:1989iv}. However, the study of the near threshold charmonium-like state $X(3872)$ indicates that the coupled channel effects are crucial for understanding the higher charmonia spectra \cite{Ferretti:2013faa, Eichten:2004uh, Pennington:2007xr}. The screened potential as a effective description of the coupled channel effect predicted a similar mass spectrum of the charmonia as the coupled channel quark model \cite{Li:2009ad, Li:2009zu}. Such a kind of screened effect is also supported by the unquenched Lattice QCD calculations \cite{Born:1989iv, Bali:2005fu} and the estimation in the light mesons \cite{Badalian:2016ttl}.

Stimulated by the similarity of the charmonium and bottomonium spectra and the anomalous mass gaps of the $S-$wave vector charmonia, i.e., if $\psi(4415)$ is assigned as $\psi(4S)$, the mass gap of $\psi(4S)$ and $\psi(3S)$ is about 382 MeV, which is larger than the one of $\psi(3S)$ and $\psi(2S)$,  the authors in Ref. \cite{He:2014xna} proposed a narrow $\psi(4S)$ near 4.2 GeV, which is consistent with the screened potential prediction \cite{Li:2009ad, Li:2009zu}. Later, the BESIII observed a new structure $X(4230)$ in the $e^+ e^- \to \chi_{c0} \omega$ \cite{Ablikim:2014qwy}, which could be a good candidate of the predicted $\psi(4S)$ \cite{Chen:2014sra, Chen:2015bma}. The prediction of $\psi(4S)$ indicates the screened potential model could provide a better description for the higher charmonia.

\begin{table}[t]
\caption{The masses of the charmonia in the $J/\psi$ and $\chi_{cJ}$ families in units of MeV. The GI, CP and CC are indicate the calculation from the Godfrey and Isgure \cite{Godfrey:1985xj}, the screened potential quark model \cite{Li:2009ad} and coupled channel quark model \cite{Pennington:2007xr}, respectively. \label{Tab:spectra}}
\begin{tabular}{ccccc}
\toprule[1pt]\toprule[1pt]
State & GI \cite{Godfrey:1985xj}  & SP \cite{Li:2009ad} & CC \cite{Pennington:2007xr} & Expt. \cite{Agashe:2014kda}\\
\midrule[1pt]
$J/\psi$    & 3098 & 3097 & 3090 & $3096.916 \pm 0.011$\\
$\psi(2S)$  & 3676 & 3673 & 3663 & $3686.093 \pm 0.034$\\
$\psi(3S)$  & 4100 & 4022 & 4036 & $4039 \pm 1$ \\
$\psi(4S)$  & 4450 & 4273 & ---  & ---
\footnote{In the GI model \cite{Godfrey:1985xj}, $\psi(4415)$ is assigned as $\psi(4S)$, while in the screened potential model, $\psi(4415)$ is considered as $\psi(5S)$.}\\
$\chi_{c0}(1P)$ & 3445 & 3433 &3415  & $3414.75 \pm 0.71$\\
$\chi_{c1}(1P)$ & 3510 & 3510 &3489& $3510.66 \pm 0.07$\\
$\chi_{c2}(1P)$ & 3550 & 3554 &3550& $3556.20 \pm 0.09$\\
$\chi_{c0}(2P)$ & 3916 & 3842 &3782 & ---\\
$\chi_{c1}(2P)$ & 3953 & 3901 &3859 &---\\
$\chi_{c2}(2P)$ & 3979 & 3937 &3917&$3929 \pm 5$\\
$\chi_{c0}(3P)$ & 4292 & 4131 &---& ---\\
$\chi_{c1}(3P)$ & 4317 & 4178 &---& ---\\
$\chi_{c2}(3P)$ & 4337 & 4208 &---&---\\
\bottomrule[1pt] \bottomrule[1pt]
\end{tabular}
\end{table}

The coupled channel quark model and the screened potential model predicted similar mass spectra of charmonia but different from the quenched quark model \cite{Li:2009ad, Li:2009zu, Pennington:2007xr, Godfrey:1985xj}. In Table \ref{Tab:spectra}, a comparison of the charmonium mass spectrum among different quark models is presented. The masses from the screened potential model and the coupled channel quark model are much lower than the quenched quark model, especially for the higher charmonia. As for $\psi(4S)$, $X(4230)$ is about 50 MeV below the screened potential model prediction, and the predicted $\chi_{c1}(3P)$ is located at 4178 MeV \cite{Li:2009ad, Li:2009zu}, which is about 40 MeV above $Y(4140)$, thus from this point of view, $Y(4140)$ could be a good candidate of the $\chi_{c1}(3P)$. With such a discrepancy of the theoretical predictions and the observation, the mass of $\chi_{c0}(3P)$ should be close to 4.09 GeV, which is also consistent with the one of the $Y(4080)$ observed in the $B\to K D\bar{D}$ process. Similarly, the mass of $\chi_{c2}(3P)$ should be about 4.17 GeV.

{\it Decay behavior:---} The open charm decays of the $\chi_{cJ}(3P)$ could be estimated in a quark pair creation (QPC) model. In this model, the quark-antiquark pair is created from the vacuum with the $J^{PC}=0^{++}$, thus the QPC model is also named $^3P_0$ model, which was proposed by Micu \cite{Micu:1968mk, LeYaouanc:1972vsx, LeYaouanc:1973ldf, LeYaouanc:1974cvx, LeYaouanc:1977fsz, LeYaouanc:1977gm} and widely used to calculate the OZI allowed decay process \cite{ Bonnaz:2001aj, Blundell:1995ev, Ackleh:1996yt, Godfrey:1986wj, Liu:2009fe, Close:2005se, Song:2014mha, Godfrey:2015dia}. For the OZI allowed strong decay process $A\to BC$, the corresponding $S-$matrix is,
\begin{eqnarray}
\langle BC \left|S \right| A\rangle = I-i (2\pi) \delta(E_f -E_i)  \langle BC \left|T \right| A\rangle.
\end{eqnarray}
In the nonrelativistic limit, the transition operator $T$ is defined as
\begin{eqnarray}
T& = &-3\gamma \sum_{m}\langle 1m;1-m|00\rangle\int d \mathbf{p}_3d\mathbf{p}_4 \delta ^3 (\mathbf{p}_3+\mathbf{p}_4) \nonumber \\
&& \times \mathcal{Y}_{1m}\left( \frac{\mathbf{p}_3-\mathbf{p}_4}{2} \right) \chi _{1,-m}^{34} \phi _{0}^{34}
\omega_{0}^{34} b_{3i}^{\dag} (\mathbf{p}_3) d_{4i}^{\dag}(\mathbf{p}_4).
\end{eqnarray}
This transition operator is introduced to describe the quark-antiquark (denoted by indices 3 and 4) created from vacuum. The phenomenological creation strength $\gamma$ for $q\bar{q}$ is taken as $\gamma=6.3$ \cite{Sun:2009tg}, while the strength for $s\bar{s}$ satisfies $\gamma_s=\gamma/\sqrt{3}$. $\mathcal{Y}_{\ell m}(\mathbf{p})=|\mathbf{p}|^\ell Y_{\ell m}({\hat{\mathbf{p}}}) $ is the $\ell$th solid harmonic polynomial. $\chi$, $\phi$, and $\omega$ are the general description of the spin, flavor, and color wave functions of the quark-antiquark pair, respectively.

The partial wave amplitudes could be related to the helicity amplitudes by \cite{Jacob:1959at}
\begin{eqnarray}
\mathcal{M}^{JL}(\mathbf{P})&=&\frac{\sqrt{2L+1}}{2J_A+1}\sum_{M_{J_B}M_{J_C}}\langle L0;JM_{J_A}|J_AM_{J_A}\rangle \nonumber \\
&&\times \langle J_BM_{J_B};J_CM_{J_C}|{J_A}M_{J_A}\rangle \mathcal{M}^{M_{J_{A}}M_{J_B}M_{J_C}},
\end{eqnarray}
and the partial width of $A\to BC$ is
\begin{eqnarray}
\Gamma_{A\to BC} &=& \pi ^2\frac{|\mathbf{P}_B|}{m_A^2}\sum_{J,L}|\mathcal{M}^{JL}(\mathbf{P})|^2,
\end{eqnarray}
where $m_{A}$ is the mass of the initial state $A$ and $\mathbf{P}_B$ is the three momentum of the $B$ in the $A$ rest frame.

In the present work, we adopt the simple harmonic oscillator (SHO) wave function $\Psi^{\mathrm{HO}}_{n,\ell m}(\mathbf{k})$ to simulate the wave functions of charmonium, charmed and charmed-strange mesons. The value of a parameter $R$ appearing in the SHO wave function can be obtained such that it reproduces the realistic root mean square (rms) radius, which can be calculated by the relativistic quark model \cite{Godfrey:1986wj}. The simple harmonic oscillator wave function with a parameter $R$ has been widely used to study the OZI allowed strong decays. In Refs. \cite{Close:2005se, Song:2014mha, Godfrey:2015dia, Barnes:2003vb}, the $R$ values of the low-lying mesons were determined by the relativistic quark model and with these $R$ parameters, the decays of the mesons were also investigated, and the obtained results were consistent with the experimental measurements. In Refs. \cite{Liu:2009fe, He:2014xna}, the decay behaviors of the higher charmonium are estimated with the simple harmonic oscillator wave functions and the obtained results could well reproduce the corresponding experimental data, which proves such an approach is reliable to investigate the strong decays of hadrons.

The unquenched relativistic quark model achieved great successes in the description of the low-lying mesons. Thus, in the present calculations, we fix the $R$ values for $D$, $D^\ast$, $D_s$, and $D_s^\ast$ to be 1.52, 1.85, 1.41 and 1.69 $\mathrm{GeV}^{-1}$, respectively, which are estimated from the relativistic quark model \cite{Godfrey:1986wj}. Different from the charmed and charmed-strange mesons, the parameter $R$ introduced by the wave function of $\chi_{cJ}(3P)$ is considered as a parameter since we are discussing the higher excited states in charmonium family, where the coupled-channel effects become important. The coupled effects not only shift the mass spectrum but also modify the wave functions of the quarkonium. In Ref. \cite{Liu:2009fe}, the width of $\chi_{c2}(2P)$ could be well reproduced when $R \sim 1.8\ \mathrm{GeV}^{-1}$. The $R$ value for the $3P$ charmonia should be a bit larger than the one of the $2P$ state, thus, we varies $R$ from 1.8 to 2.6 $\mathrm{GeV}^{-1}$ for $\chi_{cJ}(3P)$ in the present work. In addition, the constituent quark masses for charm, up/down, and strange quarks are adopted to be 1.60, 0.22, and 0.419 GeV, respectively \cite{Liu:2009fe}

\begin{figure}[t]
\scalebox{0.55}{\includegraphics{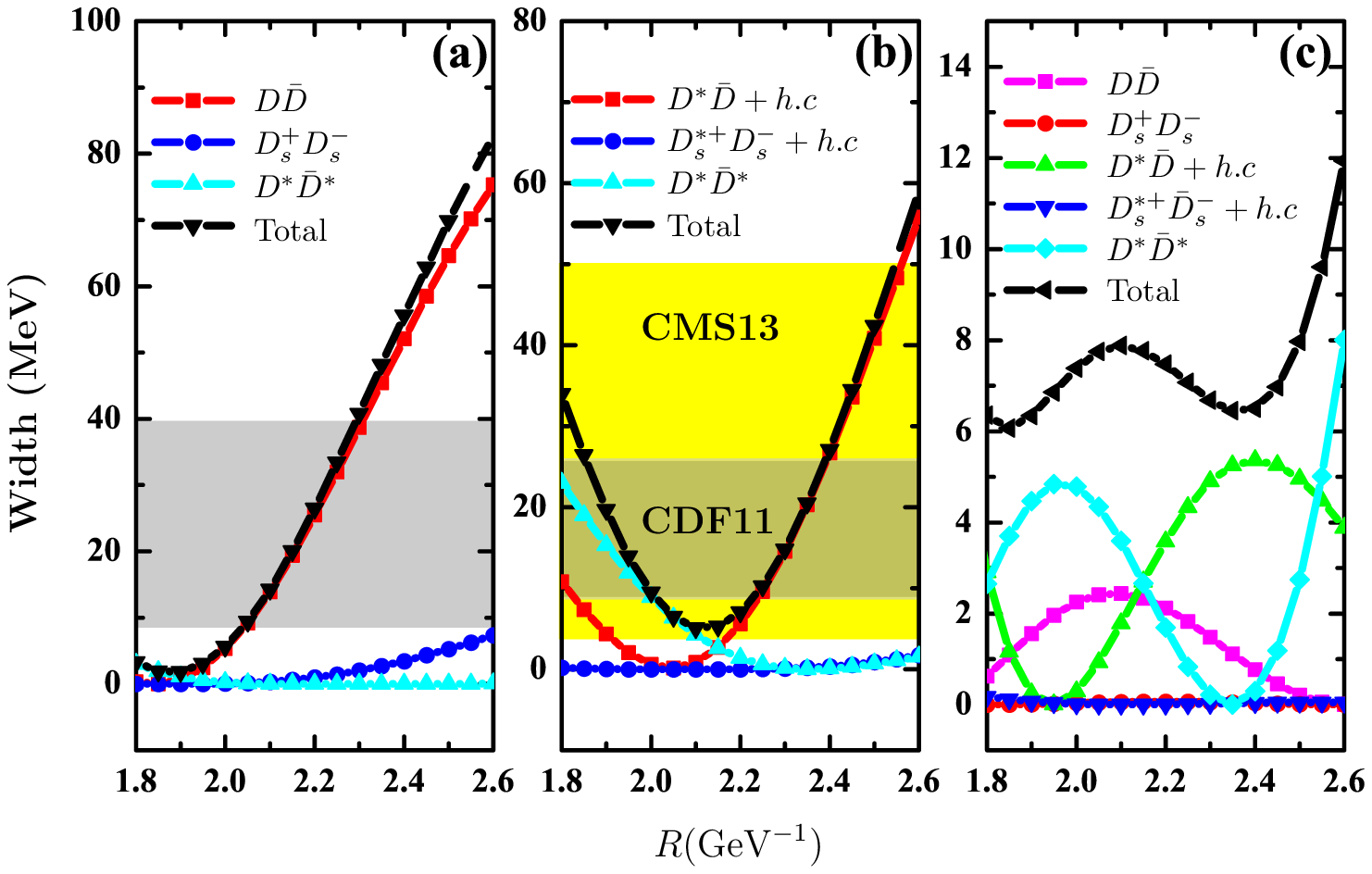}} \caption{The partial and total widths of the open charm decays of the $\chi_{cJ}(3P)$ states. $a-c$ correspond to the $J=0$, 1 and 2, respectively. The light grey band in diagram (a) is corresponding to the total width of the $\chi_{c0}(3P)$ determined in the present work. The yellow and light gray bands in diagram (b) are corresponding to the total width of the $\chi_{c1}(3P)$ reported by CDF \cite{Aaltonen:2011at} and CMS \cite{Chatrchyan:2013dma} Collaborations, respectively.  \label{Fig:decay}}
\end{figure}

The $R$ dependences of the total and partial widths of the $\chi_{cJ}(3P)$ states are presented in Fig. \ref{Fig:decay}. Based on the estimations in the present work, we find

(1) As shown in Fig. \ref{Fig:decay}b, the estimated total width of $\chi_{c1}(3P)$ could overlap with the experimental data reported by CDF \cite{Aaltonen:2011at} and CMS \cite{Chatrchyan:2013dma} Collaborations, but it can only reach up to the lower limit of the LHCb measurement \cite{Aaij:2016iza, Aaij:2016nsc}, which indicates that the present calculations support $Y(4140)$ to be a narrow state and is also consistent with our fit to the $\chi_{c1} \pi^+ \pi^-$ invariant mass spectrum of the $B\to K \chi_{c1} \pi^+ \pi^-$ process.

(2) The estimated total and partial widths of $\chi_{c0}(3P)$ are presented in Fig. \ref{Fig:decay}a. For comparison, we also present the width obtained from our fit to the $D\bar{D}$ invariant mass spectrum of the $B\to K D\bar{D}$ process. The estimated total width is consistent with the fitted one in the range of $2.05<R<2.3\ \mathrm{GeV}^{-1}$, which is similar to the one determined by the total width of $\chi_{c1}(3P)$. In this $R$ range, $\chi_{c0}(3P)$ dominantly decays into $D\bar{D}$, which is why we could observe its signal in the $D\bar{D}$ invariant mass spectra.

(3) For $\chi_{c2}(3P)$, the partial widths are strongly dependent on the parameter $R$ due to nodes in the wave function of the charmonium. However, the estimated total width of the $\chi_{c2}(3P)$ state is rather stable, which is of the order 10  MeV in the considered $R$ range. This particular property of $\chi_{c2}(3P)$ could be a crucial test to the present calculations.

To summarize, we find a narrow resonance in the $\chi_{c1} \pi^+ \pi^-$ invariant mass spectrum of the $B\to K \chi_{c1} \pi^+ \pi^-$ process with the resonance parameters consistent with the values of $Y(4140)$ reported in the $J/\psi \phi $ invariant mass spectrum, which indicates that the resonance in both processes could be the same one. $Y(4140)$ is assigned as $\chi_{c1}(3P)$ rather than a $D_s^{\ast+}   D_s^{-}$ molecule state or $c\bar{c} s\bar{s}$ tetraquark state due to its coupling to $\chi_{c1} \pi^+ \pi^-$ and the charmonium mass spectrum predicted by the screened potential model. $B\to K \chi_{c1} \pi^+ \pi^-$ could be a good process of observing $Y(4140)$ since $Y(4140)$ is far above the threshold of $\chi_{c1} \pi^+ \pi^-$, which could avoid the pollution of the threshold effect and under threshold resonance. In addition, we find a $\chi_{c0} (3P)$ candidate in the $D\bar{D}$ invariant mass spectrum of the $B\to K D\bar{D}$ process. The mass and width of this new state is fitted to be $4083.0 \pm 5.0 $ MeV and $24.1 \pm 15.4$ MeV, respectively. Our calculations of the total widths of $\chi_{c0}(3P)$ and $\chi_{c1}(3P)$ in the quark pair creation model are well consistent with those extracted from the experimental data. Furthermore, the present calculations also indicate that $\chi_{c2}(3P)$ is a rather narrow state. These results in the present work could be tested by further experiments in the LHCb and forthcoming Belle II.

Besides the open charm decays of the $\chi_{cJ}(3P)$ states, the hidden charm decay processes of these states are also important. For example, the candidate of $\chi_{c1}(3P)$, $Y(4140)$ is observed in the $J/\psi \phi$ and $\chi_{c1} \pi^+ \pi^-$ modes and, in principle, this state could also decay into $J/\psi \omega$. At present, there exist the measurements of the $J/\psi \omega$ invariant mass spectrum of $\gamma \gamma \to J/\psi \omega$ \cite{Lees:2012xs, Uehara:2009tx} and $B\to K J/\psi \omega$ \cite{Abe:2004zs, delAmoSanchez:2010jr} processes from both Belle and BaBar Collaborations. For the former process, $Y(4140)$ is forbidden with the $\chi_{c1}(3P)$ assignment. For the $B\to K J/\psi \omega$ process, one can find the bin size of the experimental data around 4.1 GeV is 40 MeV, which is about 2 to 3 times larger than the width of $Y(4140)$. Thus, It would be difficult to find any evidence of $Y(4140)$ in the present experimental data.  We expect that the future precise experimental measurements of the $B\to K J/\psi \omega$ could provide us more information of the structures in the $J/\psi \omega$ invariant mass spectrum.

Before the end of this work, we mention that the $\gamma \gamma $ collision could be an ideal process of searching $P$ wave charmonia, in which $X(3915)$ and $Z(3930)$ had been observed. In the $\gamma \gamma \to D\bar{D}$ process, there is a single bin bump at 4.085 GeV in the Belle data \cite{Uehara:2005qd}, while in the Babar data a bump appears at 4.095 GeV \cite{Aubert:2010ab}. However, since the experimental data in this energy range have large errors, more precise measurements from the forthcoming Belle II could further check the relation of this bump and the $\chi_{c0}(3P)$ state.

\section*{Acknowledgement}
The author would like to thank Estia Eichten for fruitful and inspired discussion. This project is supported by the National Natural Science Foundation of China under Grant No. 11375240.

\end{document}